\documentclass[11pt,letterpaper]{article}
\usepackage[margin=1in]{geometry}
\usepackage[ruled]{algorithm2e} 

\SetAlFnt{\small}
\SetAlCapFnt{\small}
\SetAlCapNameFnt{\small}
\SetAlCapHSkip{0pt}
\IncMargin{-\parindent}
\usepackage[square,numbers,sort]{natbib}

\usepackage{xspace}
\usepackage{setspace}
\usepackage{amsmath,amsthm,amssymb,bbm}
\usepackage{cleveref}
\crefname{algocf}{alg.}{algs.}
\Crefname{algocf}{Algorithm}{Algorithms}
\SetKwInOut{KwParam}{Parameters}

\newtheorem{theorem}{Theorem}
\newtheorem{lemma}{Lemma}

\newtheorem{proposition}{Proposition}
\theoremstyle{definition}

\renewcommand{\bar}{\overline}
\renewcommand{\hat}{\widehat}
\renewcommand{\tilde}{\widetilde}
\newcommand{\E}{\mathbb{E}}
\newcommand{\id}{\mathbbm{1}}
\newcommand{\floor}[1]{\lfloor {#1} \rfloor}
\newcommand{\ceil}[1]{\lceil {#1} \rceil}
\DeclareMathOperator*{\argmax}{arg\,max}
\DeclareMathOperator*{\argmin}{arg\,min}
\newcommand{\set}[1]{{\left\{#1\right\}}}

\newcommand{\abs}[1]{\left|#1\right|}
\newcommand{\norm}[1]{\left\|#1\right\|}
\newcommand{\bbN}{\mathbb{N}}
\newcommand{\bbR}{\mathbb{R}}
\newcommand{\bigOh}[1]{\mathcal{O}\left( #1 \right)}
\newcommand{\bigOhLog}[1]{\tilde{\mathcal{O}}\left( #1 \right)}
\newcommand{\bigOm}[1]{\Omega\left( #1 \right)}

\newcommand{\bigTh}[1]{\Theta\left( #1 \right)}

\newcount\Comments  
\usepackage{color}
\definecolor{darkgreen}{rgb}{0,0.5,0}
\definecolor{purple}{rgb}{1,0,1}
\newcommand{\kibitz}[2]{\ifnum\Comments=1\textcolor{#1}{#2}\fi}

\newcommand{\nsw}{\operatorname{NSW}}
\newcommand{\mustar}{\mu^*}
\newcommand{\muhat}{\hat{\mu}}
\newcommand{\pstar}{p^*}
\newcommand{\phat}{\hat{p}}
\newcommand{\explorefirst}{Explore-First\xspace}
\newcommand{\epsgreedy}{Epsilon-Greedy\xspace}
\newcommand{\epstgreedy}{$\epsilon^t$-Greedy\xspace}
\newcommand{\ucb}{UCB\xspace}
\newcommand{\cleanmu}{\mathcal{E}}
\newcommand{\cleanmcd}{\mathcal{H}}

\newcommand{\cleanalpha}{\mathcal{C}_{\alpha}}
\newcommand{\cleanepsgr}{\mathcal{B}}
\newcommand{\cleanucb}{\mathcal{B}}
\newcommand{\calP}{\mathcal{P}}
\newcommand{\br}[1]{\left[#1\right]}

\begin{document}\allowdisplaybreaks

\title{Fair Algorithms for Multi-Agent Multi-Armed Bandits}

\author{Safwan Hossain\\University of Toronto\\\texttt{safwan.hossain@mail.utoronto.ca}
	\and
	Evi Micha\\University of Toronto\\\texttt{emicha@cs.toronto.edu}
	\and
	Nisarg Shah\\University of Toronto\\\texttt{nisarg@cs.toronto.edu}
}

\date{}

\maketitle

\begin{abstract}
	We propose a multi-agent variant of the classical multi-armed bandit problem, in which there are $N$ agents and $K$ arms, and pulling an arm generates a (possibly different) stochastic reward for each agent. Unlike the classical multi-armed bandit problem, the goal is not to learn the ``best arm''; indeed, each agent may perceive a different arm to be the best for her personally. Instead, we seek to learn a fair distribution over the arms. Drawing on a long line of research in economics and computer science, we use the \emph{Nash social welfare} as our notion of fairness. We design multi-agent variants of three classic multi-armed bandit algorithms and show that they achieve sublinear regret, which is now measured in terms of the lost Nash social welfare.
\end{abstract}

\section{Introduction}
In the classical (stochastic) multi-armed bandit (MAB) problem, a principal has access to $K$ arms and pulling arm $j$ generates a stochastic reward for the principal from an unknown distribution $D_j$ with an unknown mean $\mustar_j$. If the mean rewards were known a priori, the principal could just repeatedly pull the \emph{best arm} given by $\argmax_j \mustar_j$. However, the principal has no apriori knowledge of the quality of the arms. Hence, she uses a learning algorithm which operates in rounds, pulls arm $j^t$ in round $t$, observes the stochastic reward generated, and uses that information to learn the best arm over time. The performance of such an algorithm is measured in terms of its cumulative regret up to a horizon $T$, defined as $\sum_{t=1}^T (\max_j \mustar_j - \mustar_{j^t})$. Note that this is the difference between the total mean reward that would have been achieved if the best arm was pulled repeatedly and the total mean reward of the arms pulled by the learning algorithm up to round $T$.

This problem can model situations where the principal is deliberating a policy decision and the arms correspond to the different alternatives she can implement. However, in many real-life scenarios, making a policy decision affects not one, but several agents. For example, imagine a company making a decision that affects all its employees, or a conference deciding the structure of its review process, which affects various research communities. This can be modeled by a multi-agent variant of the multi-armed bandit (MA-MAB) problem, in which there are $N$ agents and pulling arm $j$ generates a (possibly different) stochastic reward for each agent $i$ from an unknown distribution $D_{i,j}$ with an unknown mean $\mustar_{i,j}$.

Before pondering about learning the ``best arm'' over time, we must ask what the best arm even means in this context. Indeed, the ``best arm'' for one agent may not be the best for another. A first attempt may be to associate some ``aggregate quality'' to each arm; for example, the quality of arm $j$ may be defined as the total mean reward it gives to all agents, i.e., $\sum_i \mustar_{i,j}$. This would nicely reduce our problem to the classic multi-armed bandit problem, for which we have an armory of available solutions~\cite{Sliv19}. However, this approach suffers from the \emph{tyranny of the majority}~\cite{Moul03}. For example, imagine a scenario with ten agents, two arms, and deterministic rewards. Suppose four agents derive a reward of $1$ from the first arm but $0$ from the second, while the remaining six derive a reward of $1$ from the second arm but $0$ from the first. The aforementioned approach will deem the second arm as the best and a classical MAB algorithm will converge to \emph{repeatedly} pulling the second arm, thus unfairly treating the first four agents (a minority). A solution which treats each group in a ``proportionally fair''~\cite{Moul03} manner should ideally converge to pulling the first arm $40\%$ of the time and the second $60\%$ of the time. Alternatively, we can allow the learning algorithm to ``pull'' a probability distribution over the arms and seek an algorithm that converges to placing probability $0.4$ on the first arm and $0.6$ on the second. 

This problem of making a fair collective decision when the available alternatives --- in this case, probability distributions over the arms --- affect multiple agents is well-studied in computational social choice~\cite{BCEL+16}. The literature offers a compelling fairness notion called the \emph{Nash social welfare}, named after John Nash. According to this criterion, the fairest distribution maximizes the \emph{product} of the expected utilities (rewards) to the agents. A distribution $p$ that places probability $p_j$ on each arm $j$ gives expected utility $\sum_j p_j \cdot \mustar_{i,j}$ to agent $i$. Hence, the goal is to maximize $\nsw(p,\mustar) = \prod_{i=1}^N (\sum_{j=1}^K p_j \cdot \mustar_{i,j})$ over $p$. One can verify that this approach on the aforementioned example indeed yields probability $0.4$ on the first arm and $0.6$ on the second, as desired. It is also interesting to point out that with a single agent ($N=1$), the distribution maximizing the Nash social welfare puts probability $1$ on the best arm, thus effectively reducing the problem to the classical multi-armed bandit problem (albeit with subtle differences which we highlight in \Cref{sec:lower-bound}). 

Maximizing the Nash social welfare is often seen as a middle ground between maximizing the utilitarian social welfare (sum of utilities to the agents), which is unfair to minorities (as we observed), and maximizing the egalitarian social welfare (minimum utility to any agent), which is considered too extreme~\cite{Moul03}. The solution maximizing the Nash social welfare is also known to satisfy other qualitative fairness desiderata across a wide variety of settings~\citep{FMS18,CFS17,ABFH+20,CKMP+19,FZC17,ABM19,FGM16,BBPS+20}. For example, a folklore result shows that in our setting such a solution will always lie in \emph{the core}; we refer the reader to the work of \citet{FMS18} for a formal definition of the core as well as a short derivation of this fact using the first-order optimality condition. For further discussion on this, see \Cref{sec:related,sec:disc}.

When \emph{exactly} maximizing the Nash social welfare is not possible (either due to a lack of complete information, as in our case, or due to computational difficulty), researchers have sought to achieve approximate fairness by approximately maximizing this objective~\citep{CDGJ+17,AMGV18,Lee17,GM19,CG18,GHM18}. Following this approach in our problem, we define the (cumulative) regret of an algorithm at horizon $T$ as $\sum_{t=1}^T (\max_p \nsw(p,\mustar)-\nsw(p^t,\mustar))$, where $p^t$ is the distribution selected in round $t$. Our goal in this paper is to design algorithms whose regret is sublinear in $T$.

\subsection{Our Results}\label{sec:results}
We consider three classic algorithms for the multi-armed bandit problem: \explorefirst, \epsgreedy, and \ucb~\cite{Sliv19}. All three algorithms attempt to balance exploration (pulling arms only to learn their rewards) and exploitation (using the information learned so far to pull ``good'' arms). \explorefirst performs exploration for a number of rounds optimized as a function of $T$ followed by exploitation in the remaining rounds to achieve regret $\bigOhLog{K^{1/3} T^{2/3}}$. \epsgreedy flips a coin in each round to decide whether to perform exploration or exploitation and achieves the same regret bound. Its key advantage over \explorefirst is that it does not need to know the horizon $T$ upfront. \ucb merges exploration and exploitation to achieve a regret bound of $\bigOhLog{K^{1/2} T^{1/2}}$. Here, $\tilde{\mathcal{O}}$ hides log factors. Traditionally, the focus is on optimizing the exponent of $T$ rather than that of $K$ as the horizon $T$ is often much larger than the number of arms $K$. It is known that the dependence of UCB's regret on $T$ is optimal: no algorithm can achieve \emph{instance-independent} $o(T^{1/2})$ regret~\cite{ACFS02}.\footnote{In instance-independent bounds, the constant inside the big-Oh notation is not allowed to depend on the (unknown) distributions in the given instance. UCB also achieves an $O(\log T)$ instance-dependent regret bound, which is also known to be asymptotically optimal~\cite{LR85}. For further discussion, see \Cref{sec:disc}.}

We propose natural multi-agent variants of these three algorithms. Our variants take the Nash social welfare objective into account and select a distribution over the arms in each round instead of a single arm. For \explorefirst, we derive $\bigOhLog{N^{2/3} K^{1/3} T^{2/3}}$ regret bound, which recovers the aforementioned single-agent bound with an additional factor of $N^{2/3}$. We also show that changing a parameter of the algorithm yields a regret bound of $\bigOhLog{N^{1/3} K^{2/3} T^{2/3}}$, which offers a different tradeoff between the dependence on $N$ and $K$. For \epsgreedy, we recover the same two regret bounds, although the analysis becomes much more intricate. This is because, as mentioned above, \epsgreedy is a horizon-independent algorithm (i.e. it does not require apriori knowledge of $T$), unlike \explorefirst. For UCB, we derive $\bigOhLog{NKT^{1/2}}$ and $\bigOhLog{N^{1/2}K^{\frac{3}{2}}T^{1/2}}$ regret bounds; our dependence on $K$ worsens compared to the classical single-agent case, but importantly, we recover the same $\sqrt{T}$ dependence. Finally, we note that even for $N=1$, a learning algorithm is slightly more powerful in our setting than in the classical setting since it can choose a distribution over the arms as opposed to a deterministic arm. Nonetheless, we derive an $\Omega(\sqrt{T})$ instance-independent lower bound on the regret of any algorithm in our setting, establishing the asymptotic optimality of our UCB variant. 

Deriving these regret bounds for the multi-agent case requires overcoming two key difficulties that do not appear in the single-agent case. First, our goal is to optimize a complicated function, the Nash social welfare, rather than simply selecting the best arm. This requires a Lipschitz-continuity analysis of the Nash social welfare function and the use of new tools such as the McDiarmid's inequality which are not needed in the standard analysis. Second, the optimization is over an infinite space (the set of distributions over arms) rather than over a finite space (the set of arms). Thus, certain tricks such as a simple union bound no longer work; we use the concept of $\delta$-covering, used heavily in the Lipschitz bandit framework~\cite{KSU08}, in order to address this. 

\subsection{Related Work}\label{sec:related}
Since the multi-armed bandit problem was introduced by \citet{Thom33}, many variants of it have been proposed, such as sleeping bandit~\cite{KNS10}, contextual bandit~\cite{Wood79}, dueling bandit~\cite{YBKJ12}, Lipschitz bandit~\cite{KSU08}, etc. However, all these variants involve a single agent who is affected by the decisions. We note that other multi-agent variants of the multi-armed bandit problem have been explored recently~\cite{chakraborty2017coordinated, bargiacchi2018learning}. However, they still involve a common reward like in the classical multi-armed bandit problem. Their focus is on getting the agents to cooperate to maximize this common reward. 

Another key aspect of our framework is the focus on fairness. Recently, several papers have focused on fairness in the multi-armed bandit problem. For instance, \citet{joseph2016fairness} design a UCB variant which guarantees what they refer to as meritocratic fairness to the arms, i.e., that a worse arm is never preferred to a better arm regardless of the algorithm's confidence intervals for them. \citet{liu2017calibrated} require that similar arms be treated similarly, i.e., two arms with similar mean rewards be selected with similar probabilities. \citet{gillen2018online} focus on satisfying fairness with respect to an unknown fairness metric. Finally, \citet{PNN20} assume that there are external constraints requiring that each arm be pulled in at least a certain fraction of the rounds and design algorithms that achieve low regret subject to this constraint. All these papers seek to achieve fairness \emph{with respect to the arms}. In contrast, in our work, the arms are ``inanimate'' (e.g. policy decisions) and we seek fairness \emph{with respect to the agents}, who are separate from the arms. 

More broadly, the problem of making a fair decision given the (possibly conflicting) preferences of multiple agents is well-studied in computational social choice~\cite{BCEL+16} in a variety of contexts. For example, one can consider our problem as that of randomized voting (alternatively known as \emph{fair mixing}~\cite{ABM19}) by viewing the agents as voters and the arms are candidates. The goal is then to pick a fair lottery over the candidates given the voters' preferences. This is also a special case of other more complex models studied in the literature such as fair public decision-making~\cite{CFS17} and fair allocation of public goods~\cite{FMS18}. However, in computational social choice, voters typically have fixed preferences over the candidates. In contrast, rewards observed by the agents in our framework are stochastic. From this viewpoint, our work provides algorithms for maximizing the Nash social welfare when noisy information can be queried regarding agent preferences. 

\section{Preliminaries}\label{sec:prelim}
For $n \in \bbN$, define $[n] = \set{1,\ldots,n}$. Let $N,K \in \bbN$. In the \emph{multi-agent multi-armed bandit} (MA-MAB) problem, there is a set of \emph{agents} $[N]$ and a set of \emph{arms} $[K]$. For each agent $i \in [N]$ and arm $j \in [K]$, there is a \emph{reward distribution} $D_{i,j}$ with mean $\mustar_{i,j}$ and support $[0,1]$;\footnote{We need the support of the distribution to be non-negative and bounded, but the upper bound of $1$ is without loss of generality. All our bounds scale linearly with the upper bound on the support.} when arm $j$ is pulled, each agent $i$ observes an independent \emph{reward} sampled from $D_{i,j}$. Let us refer to $\mustar = (\mustar_{i,j})_{i \in [N],j \in [K]} \in [0,1]^{N\times K}$ as the (true) reward matrix. 

\paragraph{Policies:} As mentioned in the introduction, pulling an arm deterministically may be favorable to one agent, but disastrous to another. Hence, we are interested in \emph{probability distributions} over arms, which we refer to as \emph{policies}. The $K$-simplex, denoted $\Delta^K$, is the set of all policies. For a policy $p \in \Delta^K$, $p_j$ denotes the probability with which arm $j$ is pulled. Note that due to linearity of expectation, the expected reward to agent $i$ under policy $p$ is $\sum_{j=1}^K p_j \cdot \mustar_{i,j}$. 

\paragraph{Nash social welfare:} The Nash social welfare is defined the product of (expected) rewards to the agents. Given $\mu = (\mu_{i,j})_{i \in [N],j \in [K]}$, and policy $p \in \Delta^K$, define $\nsw(p,\mu) = \prod_{i=1}^N \left(\sum_{j=1}^K p_j \cdot \mu_{i,j}\right)$. Thus, the (true) Nash social welfare under policy $p$ is $\nsw(p,\mustar)$. Hence, if we knew $\mustar$, we would pick an \emph{optimal policy} $\pstar \in \argmax_{p \in \Delta^K} \nsw(p,\mustar)$. However, because we do not know $\mustar$ in advance, our algorithms will often produce an estimate $\muhat$, and use it to choose a policy; the quantity $\nsw(p,\muhat)$ will play a key role in our algorithms and their analysis.

\paragraph{Algorithms:} An algorithm for the MA-MAB problem chooses a policy $p^t$ in each round $t \in \bbN$. Then, an arm $a^t$ is sampled according to policy $p^t$, and for each agent $i \in [N]$, a reward $X^t_{i,a^t}$ is sampled independently from distribution $D_{i,a^t}$. At the end of round $t$, the algorithm learns the sampled arm $a^t$ and the reward vector $(X^t_{i,a^t})_{i \in [N]}$, which it can use to choose policies in the later rounds. 

\paragraph{Reward estimates:} All our algorithms maintain an estimate of the mean reward matrix $\mustar$ at every round. For round $t$ and arm $j \in [K]$, let $n_j^t = \sum_{s=1}^{t-1} \id[a^s = j]$ denote the number of times arm $j$ is pulled at the beginning of round $t$, and let $\muhat_{i,j}^t = \frac{1}{n_j^t} \sum_{s \in [t-1] : a^s = j} X^s_{i,j}$ denote the average reward experienced by agent $i$ from the $n_j^t$ pulls of arm $j$ thus far. Our algorithms treat these as an estimate of $\mustar_{i,j}$ available at the beginning of round $t$. Let $\muhat^t = (\muhat^t_{i,j})_{i \in [N],j \in [K]}$. 

\paragraph{Regret:} Recall that $\pstar$ is an optimal policy that has the highest Nash social welfare. The \emph{instantaneous regret} in round $t$ due to an algorithm choosing $p^t$ is $r^t = \nsw(\pstar,\mustar) - \nsw(p^t,\mustar)$. The (cumulative) \emph{regret} in round $T$ due to an algorithm choosing $p^1,\ldots,p^T$ is $R^T = \sum_{t=1}^T r^t$. We note that $R^T$ and $r^t$ are defined for a specific algorithm, which will be clear from the context. We are interested in bounding the \emph{expected regret} $\E[R^T]$ of an algorithm at round $T$, where the expectation is over the randomness involved in sampling the arms $a^t$ and the agent rewards $(X^t_{i,a^t})_{i \in [N]}$ for $t \in [T]$.\footnote{The algorithms we study do not introduce any further randomness in choosing the policies.} We say that an algorithm is \emph{horizon-dependent} if it needs to know $T$ in advance in order to yield bounded regret at round $T$, and \emph{horizon-independent} if it yields such a bound without knowing $T$ in advance. 

\paragraph{$\mathbf{\delta}$-Covering:} Given a metric space $(X,d)$ and $\delta > 0$, a set $S \subseteq X$ is called a $\delta$-\emph{cover} if for each $x \in X$, there exists $s \in S$ with $d(x,s) \le \delta$. That is, from each point in the metric space, there is a point in the $\delta$-cover that is no more than $\delta$ distance away. We will heavily use the fact that there exists a $\delta$-cover of $(\Delta^K,\norm{\cdot}_1)$ (i.e. the $K$-simplex under the $L_1$ distance) with size at most $\left(1+2/\delta\right)^K$~\cite[p.~126]{Wain19}, which follows from a simple discretization of the simplex.

\section{\explorefirst}\label{sec:explore-first}

\begin{algorithm}[htb!]
	\setstretch{1.2}
	\DontPrintSemicolon 
	\SetAlgoLined
	\KwIn{Number of agents $N$, number of arms $K$, horizon $T$}
	\KwParam{Exploration period $L$}
	
	\tcp{Pull each arm $L$ times}
	\For(\tcp*[f]{Exploration}){$t = 1,\ldots,K\cdot L$} {
		$j \gets \ceil{t/L}$\;
		$p^t \gets$ policy that puts probability $1$ on arm $j$ \tcp*{Pull arm $j$ deterministically}
	}
	Compute the estimated reward matrix $\muhat \triangleq \muhat^{K\cdot L+1}$ of the rewards observed so far\;
	Compute $\phat \in \argmax_{p \in \Delta^K} \nsw(p,\muhat)$\;
	\For(\tcp*[f]{Exploitation}){$t = K \cdot L+1,\ldots,T$} {
		$p^t \gets \phat$\;
	}
	\caption{\explorefirst}
	\label{algo:explore-first}
\end{algorithm}

Perhaps the simplest algorithm (with a sublinear regret bound) in the classic single-agent MAB framework is \explorefirst. It is composed of two distinct stages. The first stage is \emph{exploration}, during which the algorithm pulls each arm $L$ times. At the end of this stage, the algorithm computes the arm $\hat{a}$ with the best estimated mean reward, and in the subsequent \emph{exploitation} stage, pulls arm $\hat{a}$ in every round. The algorithm is horizon-dependent, i.e., it takes the horizon $T$ as input and sets $L$ as a function of $T$. Setting $L = \bigTh{K^{-\frac{2}{3}} T^{\frac{2}{3}} \log^{\frac{1}{3}}(T)}$ yields regret bound $\E[R^T] = \bigOh{K^{\frac{1}{3}} T^{\frac{2}{3}} \log^{\frac{1}{3}}(T)}$~\cite{Sliv19}.

In our multi-agent variant, presented as \Cref{algo:explore-first}, the exploration stage pulls each arm $L$ times as before. However, at the end of this stage, the algorithm computes, not an arm $\hat{a}$, but a policy $\phat$ with the best estimated Nash social welfare. During exploitation, it then uses policy $\phat$ in every round. With an almost identical analysis as in the single-agent setting, we recover the aforementioned regret bound with an additional $N^{2/3}$ factor for $N$ agents. 

Using a novel and more intricate argument, we show that a different tradeoff between the exponents of $N$ and $K$ can be obtained, where $N^{2/3}$ is reduced to $N^{1/3}$ at the expense of increasing $K^{1/3}$ to $K^{2/3}$ (and adding a logarithmic term). We later use this approach again in our analysis of more sophisticated algorithms. 

Before we proceed to the proof, we remark that \Cref{algo:explore-first} can be implemented efficiently. The only non-trivial step is to compute the optimal policy given an estimated reward matrix, i.e., $\hat{p} \in \argmax_{p \in \Delta^K} \nsw(p,\muhat)$. Since the Nash social welfare is known to be log-concave, this can be solved in polynomial time~\cite{eisenberg1959consensus}. 

We begin by presenting a few elementary lemmas regarding the behavior of the Nash social welfare function $\nsw(p,\mu)$. We are mainly interested in how much the function can change when its arguments change. To that end, the following folklore result translates the difference in a product to a sum of point-wise differences that are easier to deal with. 

\begin{lemma}\label{lem:ai-bi}
	Let $a_i,b_i \in [0,1]$ for $i \in [N]$. Then, $\abs{\prod_{i=1}^N a_i - \prod_{i=1}^N b_i} \le \sum_{i=1}^N \abs{a_i-b_i}$.	
\end{lemma}
\begin{proof}
	We prove this using induction on $N$. For $N=1$, the lemma trivially holds. Suppose it holds for $N=n$. For $N=n+1$, we have
	\begin{align*}
	\abs{\prod_{i=1}^{n+1} a_i - \prod_{i=1}^{n+1} b_i} &= \abs{\prod_{i=1}^{n+1} a_i - b_{n+1} \prod_{i=1}^n a_i + b_{n+1} \prod_{i=1}^n a_i -\prod_{i=1}^{n+1} b_i}\\
	&\le \left(\prod_{i=1}^n a_i\right) \abs{a_{n+1}-b_{n+1}} + b_{n+1} \cdot \abs{\prod_{i=1}^n a_i - \prod_{i=1}^n b_i}\\
	&\le \abs{a_{n+1}-b_{n+1}} + \sum_{i=1}^n \abs{a_i-b_i} = \sum_{i=1}^{n+1} \abs{a_i-b_i},
	\end{align*}
	where the second transition is due to the triangle inequality, and the third transition holds due to the induction hypothesis and because $a_i,b_i \in [0,1]$ for each $i$. 
\end{proof}

Using \Cref{lem:ai-bi}, we can easily analyze Lipschitz-continuity of $\nsw(p,\mu)$ when either $p$ or $\mu$ changes and the other is fixed. First, we consider change in $p$ with $\mu$ fixed. 

\begin{lemma}\label{lem:lipschitz-p}
	Given a reward matrix $\mu \in [0,1]^{N\times K}$ and policies $p^1,p^2 \in \Delta^K$, we have 
	\[
	\abs{\nsw(p^1,\mu)-\nsw(p^2,\mu)} \le N \cdot \norm{p^1-p^2}_1 = N\cdot \sum_{j \in [K]} \abs{p^1_j-p^2_j}.
	\] 
\end{lemma}
\begin{proof}
	Using \Cref{lem:ai-bi}, we have 
	\[
	\textstyle\abs{\nsw(p^1,\mu)-\nsw(p^2,\mu)} \le \sum_{i \in [N]} \abs{\sum_{j \in [K]} (p^1_j-p^2_j) \cdot \mu_{i,j}} \le N \cdot \sum_{j \in [K]} \abs{p^1_j-p^2_j},
	\]
	where the final transition is due to the triangle inequality and because $\mu_{i,j} \in [0,1]$ for each $i,j$.  
\end{proof}

Next, we consider change in $\mu$ with $p$ fixed. 

\begin{lemma}\label{lem:lipschitz-mu}
	Given a policy $p \in \Delta^K$, and reward matrices $\mu^1,\mu^2 \in [0,1]^{N\times K}$, we have 
	\[
	\abs{\nsw(p,\mu^1)-\nsw(p,\mu^2)} \le \sum_{i \in [N]} \sum_{j \in [K]} p_j \cdot \abs{\mu^1_{i,j}-\mu^2_{i,j}}.
	\]
\end{lemma}
\begin{proof}
	Again, using \Cref{lem:ai-bi}, we have
	\[
	\textstyle\abs{\nsw(p,\mu^1)-\nsw(p,\mu^2)} \le \sum_{i \in [N]} \abs{\sum_{j \in [K]} p_j \cdot (\mu^1_{i,j}-\mu^2_{i,j})} \le \sum_{i \in [N], j \in [K]} p_j \cdot \abs{\mu^1_{i,j}-\mu^2_{i,j}},
	\]
	where the last transition is due to the triangle inequality.
\end{proof}

We are now ready to derive the regret bounds for \explorefirst.

\begin{theorem}\label{thm:explore-first}
	\explorefirst is horizon-dependent and has the following expected regret at round $T$.
	\begin{itemize}
		\item  When $L = \bigTh{N^{\frac{2}{3}} K^{-\frac{2}{3}} T^{\frac{2}{3}} \log^{\frac{1}{3}}(N K T)}$, $\E[R^T] = \bigOh{N^{\frac{2}{3}} K^{\frac{1}{3}} T^{\frac{2}{3}} \log^{\frac{1}{3}}(N K T)}$.
		\item  When $L = \bigTh{N^{\frac{1}{3}} K^{-\frac{1}{3}} T^{\frac{2}{3}} \log^{\frac{2}{3}}(N K T)}$, $\E[R^T] = \bigOh{N^{\frac{1}{3}} K^{\frac{2}{3}} T^{\frac{2}{3}} \log^{\frac{2}{3}}(N K T)}$.
	\end{itemize}
\end{theorem}
\begin{proof}
	Note that the instantaneous regret $r^t(p^t)$ in any round $t$ can be at most $1$ because $\nsw(p,\mustar) \in [0,1]$ for every policy $p$. Thus, 
	\begin{equation}\label{eqn:explore-first-sketch}
	\E[R^T] = \sum_{t=1}^T \E[r^t] \le K L \cdot 1 + (T- K L) \cdot \E[\nsw(\pstar,\mustar)-\nsw(\phat,\mustar)].
	\end{equation}
	Thus, our goal is to bound $\E[\nsw(\pstar,\mustar)-\nsw(\phat,\mustar)]$. We bound this in two ways. 
	
	\medskip\noindent\textbf{First approach:} We present this approach briefly since it largely mimics the classical analysis with an application of \Cref{lem:lipschitz-mu}. Here, we bound how much $\muhat$ can deviate from $\mustar$. Specifically, we let $\epsilon = \sqrt{\frac{\log(N K T)}{L}}$ and define the event $\cleanmu \triangleq \forall i \in [N], \forall j \in [K] : \abs{\muhat_{i,j}-\mustar_{i,j}} \le \epsilon$. Since $L$ is fixed, we have $\E[\muhat_{i,j}] = \mustar_{i,j}$. Hence, we can apply Hoeffding's inequality followed by the union bound to derive $\Pr[\cleanmu] \ge 1-2/T^2$. Conditioned on $\cleanmu$, from \Cref{lem:lipschitz-mu} we have $\nsw(p,\mustar)-\nsw(p,\muhat) \le N \epsilon$ for every policy $p$, which implies 
	\begin{align*}
	\nsw(\pstar,\mustar) \le \nsw(\pstar,\muhat) + N \epsilon \le \nsw(\phat,\muhat) + N\epsilon \le \nsw(\phat,\mustar) + 2N\epsilon,
	\end{align*}
	where the second transition is because $\phat \in \argmax_{p \in \Delta^K} \nsw(p,\muhat)$. Substituting this into \Cref{eqn:explore-first-sketch}, using the fact that $\E[\nsw(\pstar,\mustar)-\nsw(\phat,\mustar)] \le 1 \cdot \E[\nsw(\pstar,\mustar)-\nsw(\phat,\mustar) | \cleanmu] + \Pr[\neg\cleanmu] \cdot 1$, and setting $L = \bigTh{N^{\frac{2}{3}} K^{-\frac{2}{3}} T^{\frac{2}{3}} \log^{\frac{1}{3}}(N K T)}$ yields the first regret bound. 
	
	\medskip\noindent\textbf{Second approach:} We now focus on another approach for bounding $\E[\nsw(\pstar,\mustar)-\nsw(\phat,\mustar)]$, which is more intricate and offers a different tradeoff between the dependence on $N$ and $K$. Notice that for a given $p$, $\E[\nsw(p,\muhat)] = \nsw(p,\mustar)$ because all $\muhat_{i,j}$-s are independent and expectation decomposes over sums and products of independent random variables. Thus, we can use McDiarmid's inequality to bound $\abs{\nsw(p,\muhat)-\nsw(p,\mustar)}$ at a given $p$. 
	
	Fix a $\delta$-cover $\calP$ of $(\Delta^K,\norm{\cdot}_1)$ with $|\calP| \le (1+2/\delta)^K$. Fix $p \in \calP$. Notice that $\muhat_{i,j} = (1/L) \cdot \sum_{s=1}^L X^s_{i,j}$, where $X^s_{i,j}$ is the reward to agent $i$ from the $s$-th pull of arm $j$ during the exploration phase.
	
	We thus decompose $\muhat$ into $N \cdot L$ random variables: for each $i \in [N]$ and $s \in [L]$, we let $X^s_i = (X^s_{i,j})_{j \in [K]}$. To apply McDiarmid's inequality, we need to analyze the maximum amount $c^s_i$ by which changing $X^s_i$ can change $\nsw(p,\muhat)$. Using \Cref{lem:lipschitz-mu}, it is easy to see that $c^s_i \le 1/L$ for each $i \in [N]$ and $s \in [L]$. Now, applying McDiarmid's inequality, we have
	\[
	\Pr\br{\abs{\nsw(p,\muhat)-\nsw(p,\mustar)} \le \epsilon} \le 2e^{\frac{-2\epsilon^2}{\sum_{i \in [N], s \in [L]} (c^s_i)^2}} = 2e^{\frac{-2 L \epsilon^2}{N}}.
	\]
	
	Setting $\epsilon = \sqrt{\frac{N \log(|\calP| T)}{2L}}$, we have that for each $p \in \calP$,  
	\[
	\Pr\br{\abs{\nsw(p,\muhat)-\nsw(p,\mustar)} \le \sqrt{\frac{N \log(|\calP| T)}{2L}}} \le \frac{2}{|\calP| T}.
	\]
	
	Using the union bound, we have that 
	\[
	\Pr\br{\forall p \in \calP: \abs{\nsw(p,\muhat)-\nsw(p,\mustar)} \le \sqrt{\frac{N \log(|\calP| T)}{2L}}} \ge 1-\frac{2}{T}.
	\]
	
	For $p \in \Delta^K$, let $\bar{p} \in \argmin_{p' \in \calP} \norm{p-p'}_1$. Then, since $\calP$ is a $\delta$-cover, we have $\norm{p-\bar{p}}_1 \le \delta$. Thus, due to \Cref{lem:lipschitz-p}, we have
	\begin{align*}
	\abs{\nsw(p,\muhat)-\nsw(p,\mustar)} &\le \sum_{\mu \in \set{\muhat,\mustar}} \abs{\nsw(p,\mu)-\nsw(\bar{p},\mu)} + \abs{\nsw(\bar{p},\muhat)-\nsw(\bar{p},\mustar)} \\
	&\le 2 N \delta + \abs{\nsw(\bar{p},\muhat)-\nsw(\bar{p},\mustar)}.
	\end{align*}
	
	Setting $\delta = \frac{1}{N T}$, we have 
	\[
	\Pr\br{\forall p \in \Delta^K: \abs{\nsw(p,\muhat)-\nsw(p,\mustar)} \le \frac{2}{T} + \sqrt{\frac{N \log(|\calP| T)}{2L}}} \ge 1-\frac{2}{T}.
	\]
	
	Next, we use the fact that 
	\[
	\nsw(\pstar,\mustar)-\nsw(\phat,\mustar) \le \sum_{p \in \set{\pstar,\phat}} \abs{\nsw(p,\muhat)-\nsw(p,\mustar)}. 
	\]
	
	Hence,
	\[
	\Pr\br{\abs{\nsw(\pstar,\mustar)-\nsw(\phat,\mustar)} \le \frac{4}{T} + \sqrt{\frac{2 N \log(|\calP| T)}{L}}} \ge 1-\frac{2}{T}.
	\]
	
	Next, we substitute $|\calP| \le (1+2/\delta)^K \le (3/\delta)^K$, $\delta = \frac{1}{N T}$, and $L = \bigTh{N^{\frac{1}{3}} K^{-\frac{1}{3}} T^{\frac{2}{3}} \log^{\frac{2}{3}}(N K T)}$, and then substitute the derived bound in \Cref{eqn:explore-first-sketch} to get the second regret bound. 
\end{proof}

\section{\epsgreedy}\label{sec:epsilon-greedy}

\newcommand{\curarm}{curr}
\begin{algorithm}[htb]
	\setstretch{1.2}
	\DontPrintSemicolon 
	\SetAlgoLined
	\KwIn{Number of agents $N$, number of arms $K$}
	\KwParam{Exploration probabilities $\epsilon^t$ for $t \in \bbN$}
	$\curarm \gets 1$ \tcp*{Next arm to pull during exploration}
	\For{$t = 1,2,\ldots,$} {
		Toss a coin with success probability $\epsilon^t$\;
		\uIf(\tcp*[f]{Exploration}){success}{
			\tcp{Round-robin among arms during exploration}
			$p^t \gets$ policy that puts probability $1$ on arm $\curarm$ \tcp*{Pull it deterministically}
			$\curarm \gets \curarm+1$ \tcp*{When $\curarm$ becomes $K+1$, reset to $1$}
		}
		\Else(\tcp*[f]{Exploitation}){
			Compute the estimated reward matrix $\muhat^t$ from the rewards observed so far\;
			$p^t \gets \argmax_{p \in \Delta^K} \nsw(p,\muhat^t)$\;
		}
	}
	\caption{\epstgreedy}
	\label{algo:epsilon-greedy}
\end{algorithm}

A slightly more sophisticated algorithm than \explorefirst is \epsgreedy, which is presented as \Cref{algo:epsilon-greedy}. It spreads out exploration instead of performing it all at the beginning. Specifically, at each round $t$, it performs exploration with probability $\epsilon^t$, and exploitation otherwise. Exploration cycles through the arms in a round-robin fashion, while exploitation uses the policy $p^t$ with the highest Nash social welfare under the current estimated reward matrix (rather than choosing a single estimated best arm as in the classical algorithm). 

We remark that, like \explorefirst, \epsgreedy can also be implemented efficiently. The only non-trivial step is to compute $\hat{p} \in \argmax_{p \in \Delta^K} \nsw(p,\muhat)$, which, as we mentioned before, can be done in polynomial time. 

The key advantage of \epsgreedy over \explorefirst is that it is horizon-independent. However, in the $\muhat$ computed in \explorefirst at the end of exploration, each $\muhat_{i,j}$ is the average of $L$ iid samples, where $L$ is fixed. In contrast, in the $\muhat^t$ computed in \epsgreedy in round $t$, each $\muhat^t_{i,j}$ is the average of $n_j^t$ iid samples. The fact that $n_j^t$ is itself a random variable and the $\muhat^t_{i,j}$-s are correlated through the $n_j^t$-s prevents a direct application of certain statistical inequalities, thus complicating the analysis of \epsgreedy. To address this, we first present a sequence of useful lemmas that apply to \emph{any} algorithm, and then use them to prove the regret bounds of \epsgreedy and later \ucb. 

\subsection{Useful Lemmas}\label{sec:lemmas}

Recall that $\mustar$ and $\muhat^t$ denote the true reward matrix and the estimated reward matrix at the beginning of round $t$, respectively. Our goal is to find an upper bound on the quantity $\abs{\nsw(p,\mustar)-\nsw(p,\muhat^t)}$ that, with high probability, holds at every $p \in \Delta^K$ simultaneously. To that end, we first need to show that $\muhat^t$ will be close to $\mustar$ with high probability.

Recall that random variable $n_j^t$ denotes the number of times arm $j$ is pulled by an algorithm before round $t$, and $\muhat^t_{i,j}$ is an average over $n_j^t$ independent samples. Hence, we cannot directly apply Hoeffding's inequality, but we can nonetheless use standard tricks from the literature. 

\begin{lemma}\label{lem:muhat-close}
	Define $r_j^t = \sqrt{\frac{2 \log (N K t)}{n_j^t}}$, and event 
	\[
	\cleanmu^t \triangleq \forall i \in [N], j \in [K] : \abs{\muhat^t_{i,j}-\mustar_{i,j}} \le r_j^t.
	\]
	Then, for any algorithm and any $t$, we have $\Pr[\cleanmu^t] \ge 1-\frac{2}{t^3}$. 
\end{lemma}
\begin{proof}
	Fix $t$. For $i \in [N]$, $j \in [K]$, and $\ell \in [t]$, let $\bar{v}^{\ell}_{i,j}$ denote the average reward to agent $i$ from the first $\ell$ pulls of arm $j$, and define $\bar{r}^{\ell}_j = \sqrt{\frac{2 \log (N K t)}{\ell}}$.  Then, by Hoeffding's inequality, we have
	\begin{align*}
	\forall i \in [N], j \in [K], \ell \in [t] : \Pr\left[\abs{\bar{v}_{i,j}^{\ell}-\mu_{i,j}} > \bar{r}_j^{\ell}\right] \le \frac{2}{(N K t)^4}.
	\end{align*}
	By the union bound, we get
	\begin{align*}
	\Pr\left[\forall i \in [N], j \in [K], \ell \in [t] :  \abs{\bar{v}_{i,j}^{\ell}-\mu_{i,j}} \le \bar{r}_j^{\ell}\right] \ge 1-\frac{2}{(N K t)^3}.
	\end{align*}
	Because $n_j^t \in [t]$ for each $j \in [K]$, the above event implies our desired event $\cleanmu^t$. Hence, we have that $\Pr[\cleanmu^t] \ge 1-2/(NKt)^3 \ge 1-2/t^3$. 
\end{proof}

Conditioned on $\cleanmu^t$, we wish to bound $\abs{\nsw(p,\mustar)-\nsw(p,\muhat^t)}$ simultaneously at all $p \in \Delta^K$. We provide two such (incomparable) bounds, which will form the crux of our regret bound analysis. The first bound is a direct application of the Lipschitz-continuity analysis from \Cref{lem:lipschitz-mu}. 

\begin{lemma}\label{lem:first-bound}
	Conditioned on $\cleanmu^t$, we have that 
	\[
	\forall p \in \Delta^K: \abs{\nsw(p,\muhat^t)-\nsw(p,\mustar)} \le N \cdot \sum_{j \in [K]} p_j \cdot r_j^t. 
	\]
\end{lemma}
\begin{proof}
	Conditioned on $\cleanmu^t$, we have $\abs{\muhat^t_{i,j}-\mustar_{i,j}} \le r_j^t$ for each $j \in [K]$. In that case, it is easy to see that the upper bound from \Cref{lem:lipschitz-mu} becomes $N \cdot \sum_{j \in [K]} p_j \cdot r_j^t$. 
\end{proof}

The factor of $N$ in \Cref{lem:first-bound} stems from analyzing how much $\muhat^t$ may deviate from $\mustar$ conditioned on $\cleanmu^t$, in the worst case. However, even after conditioning on $\cleanmu^t$, $\muhat^t$ remains a random variable. Hence, one may expect that its deviation, and thus the difference $\abs{\nsw(p,\muhat^t)-\nsw(p,\mustar)}$, may be smaller in expectation. Thus, to derive a different bound than in \Cref{lem:first-bound}, we wish to apply McDiarmid's inequality. However, there are two issues in doing so directly.

\begin{itemize}
	\item McDiarmid's inequality bounds the deviation of $\nsw(p,\muhat^t)$ from its expected value. If $\muhat^t$ consisted of independent random variables, like in \explorefirst, this would be equal to $\nsw(p,\mustar)$. However, in general, these variables may be correlated through $n_j^t$. We use a conditioning trick to address this issue. 
	\item We cannot hope to apply McDiarmid's inequality at each $p \in \Delta^K$ separately and use the union bound because $\Delta^K$ is infinite. So we apply it at each $p$ in a $\delta$-cover of $\Delta^K$, apply the union bound, and then translate the guarantee to nearby $p \in \Delta^K$ using the Lipschitz-continuity analysis from \Cref{lem:lipschitz-p}. 
\end{itemize}

The next result is one of the key technical contributions of our work with a rather long proof.

\begin{lemma}\label{lem:second-bound}
	Define the event 
	\[
	\cleanmcd^t \triangleq \forall p \in \Delta^K : \abs{\nsw(p,\muhat^t)-\nsw(p,\mustar)} \le \sqrt{12 N K  \log(N K t)} \cdot \sum_{j \in [K]} p_j \cdot r_j^t +\frac{4}{t}. 
	\]
	Then, for any algorithm and any $t$, we have $\Pr[\cleanmcd^t | \cleanmu^t] \ge 1-2/t^3$. 
\end{lemma}
\begin{proof}
Fix $p \in \Delta^K$. Fix $\delta > 0$, and let $\calP$ be a $\delta$-cover of the policy simplex $\Delta^K$ with $|\calP| \le \left(1+2/\delta\right)^K$~\cite[p.~126]{Wain19}. 
	
Conditioned on $\cleanmu^t$ (i.e. $\abs{\muhat^t_{i,j}-\mustar_{i,j}} \le r_j^t = \sqrt{\frac{2 \log(N K t)}{n_j^t}}, \forall i \in [N],j \in [K]$), we wish to derive a high probability bound on $\abs{\nsw(p,\muhat^t)-\nsw(p,\mustar)}$. We can bound the deviation of $\nsw(p,\muhat^t)$ from its expected value. However, unlike in the case of \explorefirst, we cannot directly claim that the expected value is $\nsw(p,\mustar)$ because, as we mentioned above, $\muhat^t$ consists of random variables that may be correlated through the random varaible $n^t = (n_1^t,\ldots,n_K^t)$ taking values in $[t]^K$. Thus, we need a more careful argument. 
	
For $i \in [N]$, $j \in [K]$, and $\ell_j \in [t]$, let $\bar{v}^{\ell_j}_{i,j}$ denote the average reward to agent $i$ from the first $\ell_j$ pulls of arm $j$, and define $\bar{r}^{\ell_j}_j = \sqrt{\frac{2\log (N K t)}{\ell_j}}$.  Let   $\ell = (\ell_1,\ldots,\ell_K) \in [t]^K$ and $\bar{v}^{\ell} = (\bar{v}^{\ell_j}_{i,j})_{i \in [N],j \in [K]}$. Each $\bar{v}^{\ell_j}_{i,j}$ is independent and satisfies $\E[\bar{v}^{\ell_j}_{i,j}] = \mustar_{i,j}$. Since expectation decomposes over sums and products of independent random variables, we have $\E[\nsw(p,\bar{v}^{\ell})] = \nsw(p,\mustar)$. 

\medskip\noindent\textbf{Evaluating conditional expectation:} 
We next argue that further conditioning on the high probability event $\cleanmu^t$ does not change the expectation by much. Formally,
	\begin{align}
	&\abs{\nsw(p,\mustar)-\E[\nsw(p,\bar{v}^{\ell})  | \cleanmu^t ]}\nonumber\\
	&= \abs{\E[\nsw(p,\bar{v}^{\ell})]  -\E\br{\nsw(p,\bar{v}^{\ell}) | \cleanmu^t }}\nonumber\\
	&= \Pr[\neg \cleanmu^t] \cdot \abs{\E\br{\nsw(p,\bar{v}^{\ell}) | \neg \cleanmu^t }-\E\br{\nsw(p,\bar{v}^{\ell}) | \cleanmu^t }}\nonumber\\
	&\le \Pr[\neg \cleanmu^t] \le \frac{2}{t^3} \le \frac{2}{t},\label{eqn:mcd-diff}
	\end{align}
	where the penultimate transition holds because $\nsw$ is bounded in $[0,1]$, and the final transition is due to \Cref{lem:muhat-close}. 
	
\medskip\noindent\textbf{Applying McDiarmid's inequality:} We first decompose $\bar{v}^{\ell}$ into $N$ random variables: for each $i \in [N]$, let $\bar{v}^{\ell}_i = (\bar{v}^{\ell}_{i,j})_{j \in [K]}$. To apply McDiarmid's inequality, we need to analyze the maximum amount $c_i$ by which changing $\bar{v}^{\ell}_i$ can change $\nsw(p,\bar{v}^{\ell})$. Fix $i \in [N]$, and fix all the variables except $\bar{v}^{\ell}_i$. Conditioned on $\cleanmu^t$, each $\bar{v}^{\ell}_{i,j}$ can change by at most $2 \bar{r}_j^{\ell_j}$. Hence, using \Cref{lem:lipschitz-mu}, we have that $c_i \le 2 \sum_{j \in [K]} p_j \cdot \bar{r}_j^{\ell_j}$. Now, applying McDiarmid's inequality, we have
\[
\	\forall \ell \in [t]^K :	\Pr\br{\abs{\nsw(p,\bar{v}^{\ell})-\E\br{\nsw(p,\bar{v}^{\ell})|\cleanmu^t }} \ge \epsilon \ |\ \cleanmu^t } \le 2e^{\frac{-2\epsilon^2}{\sum_{i \in [N]} c_i^2}} \le 2e^{\frac{-2\epsilon^2}{4 N \cdot \left(\sum_{j \in [K]} p_j \cdot \bar{r}_j^{\ell_j}\right)^2}}.
\]
	
Using \Cref{eqn:mcd-diff}, and setting $\epsilon = \sqrt{2 N \log(|\calP| t^{K+3})} \cdot \sum_{j \in [K]} p_j \cdot \bar{r}_j^{\ell_j}$, we have that 
\[
\forall \ell \in [t]^K :	\Pr\br{\abs{\nsw(p,\bar{v}^{\ell})-\nsw(p,\mustar)} \ge \sqrt{2 N \log(|\calP|  t^{K+3})} \cdot {\sum_{j \in [K]}} p_j \cdot \bar{r}_j^{\ell_j} +\frac{2}{t} \ \Big|\ \cleanmu^t } \le \frac{2}{|\calP| t^{K+3}}.
\]
Next, by union bound, we get	
\[
\Pr\br{	\forall \ell \in [t]^K :\abs{\nsw(p,\bar{v}^{\ell})-\nsw(p,\mustar)} \ge \sqrt{2 N \log(|\calP|  t^{K+3})} \cdot {\sum_{j \in [K]}} p_j \cdot \bar{r}_j^{\ell_j} +\frac{2}{t} \ \Big|\ \cleanmu^t } \le \frac{2}{|\calP| t^3}.
\]
Because $n_j^t \in [t]$ for each $j \in [K]$, we have 
\[
\Pr\br{	\abs{\nsw(p,\muhat^t)-\nsw(p,\mustar)} \ge \sqrt{2 N \log(|\calP|  t^{K+3})} \cdot {\sum_{j \in [K]}} p_j \cdot r_j^{t} +\frac{2}{t} \ \Big|\ \cleanmu^t } \le \frac{2}{|\calP| t^3}.
\]
	
\medskip\noindent\textbf{Extending to all policies in $\mathbf{\calP}$:} Using the union bound, we have that 
\[
\Pr\br{\forall p \in \calP: \abs{\nsw(p,\muhat^t)-\nsw(p,\mustar)} \le \sqrt{2 N \log(|\calP|  t^{K+3})} \cdot \sum_{j \in [K]} p_j \cdot r_j^t +\frac{2}{t}\ \bigg|\ \cleanmu^t} \ge 1-\frac{2}{t^3}.
\]
	
\medskip\noindent\textbf{Extending to all policies in $\mathbf{\Delta^K}$:} For $p \in \Delta^K$, let $\bar{p} \in \argmin_{p' \in \calP} \norm{p-p'}_1$. Then, since $\calP$ is a $\delta$-cover, we have $\norm{p-\bar{p}}_1 \le \delta$. Thus, due to \Cref{lem:lipschitz-p}, we have
\begin{align*}
\abs{\nsw(p,\muhat^t)-\nsw(p,\mustar)} &\le \sum_{\mu \in \set{\muhat^t,\mustar}} \abs{\nsw(p,\mu)-\nsw(\bar{p},\mu)} \\
&\qquad + \abs{\nsw(\bar{p},\muhat^t)-\nsw(\bar{p},\mustar)} \\
&\le 2 N \delta + \abs{\nsw(\bar{p},\muhat^t)-\nsw(\bar{p},\mustar)}.
\end{align*}

Setting $\delta = \frac{1}{N t}$, we have 
\[
\Pr\br{\forall p \in \Delta^K: \abs{\nsw(p,\muhat^t)-\nsw(p,\mustar)} \le  \sqrt{2 N \log(|\calP|  t^{K+3})} \cdot \sum_{j \in [K]} p_j \cdot r_j^t +\frac{4}{t}\ \bigg|\ \cleanmu^t} \ge 1-\frac{2}{t^3}.
\]
Substituting $|\calP| \le (1+2/\delta)^K \le (3/\delta)^K$ with $\delta = \frac{1}{N t}$ yields the desired bound. 
\end{proof}

Finally, we use the following simple observation in deriving our asymptotic bounds. 
\begin{proposition}\label{lem:powers}
	For constant $p \in \bbR$, $\sum_{t=1}^T t^p$ is $\bigTh{\log T}$ if $p = -1$ and $\bigTh{T^{p+1}}$ otherwise.
\end{proposition}

\subsection{Analysis of \epsgreedy}\label{sec:epsgreedy-analysis}

We can now use these lemmas to derive the regret bounds for \epsgreedy.

\begin{theorem}\label{thm:epsgreedy}
	\epsgreedy is horizon-independent, and has the following expected regret at any round $T$.
	\begin{itemize}
		\item  If $\epsilon^t = \bigTh{N^{\frac{2}{3}} K^{\frac{1}{3}} t^{-\frac{1}{3}} \log^{\frac{1}{3}}(N K t)}$ for all $t$, $\E[R^T] = \bigOh{N^{\frac{2}{3}} K^{\frac{1}{3}} T^{\frac{2}{3}} \log^{\frac{1}{3}}(N K T)}$.
		\item  If $\epsilon^t = \bigTh{N^{\frac{1}{3}} K^{\frac{2}{3}} t^{-\frac{1}{3}} \log^{\frac{2}{3}}(N K t)}$ for all $t$, $\E[R^T] = \bigOh{N^{\frac{1}{3}} K^{\frac{2}{3}} T^{\frac{2}{3}} \log^{\frac{2}{3}}(N K T)}$.
	\end{itemize}
\end{theorem}
\begin{proof}
	Fix $t \in [T]$. Let $b^t$ denote the number of times \epsgreedy performs exploration up to round $t$. Note that $\E[b^t] = \sum_{s=1}^t \epsilon^s \ge t \epsilon^t$, where the last step follows from the fact that $\epsilon^t$ is monotonically decreasing in both cases of the theorem. Let $\theta > 0$ be a constant such that $\epsilon^t \ge \theta \cdot t^{-1/3}$ in both cases of the theorem. 
	
	Define the event $\cleanepsgr^t \triangleq b^t \ge \gamma \cdot t \epsilon^t$, where $\gamma = 1-1/\theta$. Then, by Hoeffding's inequality, we have
	\begin{equation}\label{eqn:cleanepsgr}
	\Pr[\neg\cleanepsgr^t] \le e^{-2(1-\gamma)^2 \theta^2 t^{1/3}} = e^{-2 t^{1/3}} \le e^{-\log t} = \frac{1}{t}.
	\end{equation}
	
	Because the algorithm performs round-robin during exploration, conditioned on $\cleanepsgr^t$, we have that $n_j^t \ge \frac{b^t}{K} \ge \frac{\gamma \cdot t \epsilon^t}{K}$ for each arm $j$,\footnote{Technically, $n_j^t \ge \floor{\frac{b^t}{K}}$ for each arm $j$, but we omit the floor for the ease of presentation.} which implies $r_j^t \le \sqrt{\frac{2 K \log(N K t)}{\gamma \cdot t \epsilon^t}}$ for each $j$. Thus, conditioned on $\cleanepsgr^t$, we have
	\begin{equation}\label{eqn:epsgreedy-bound}
	\forall p \in \Delta^K : \sum_{j \in [K]} p_j \cdot r_j^t \le \max_{j \in [K]} r_j^t \le \sqrt{\frac{2 K \log(N K t)}{\gamma \cdot t \epsilon^t}}.
	\end{equation}
	
	We are now ready to use the bounds from \Cref{lem:first-bound,lem:second-bound}. We focus on the event 
	\[
	\cleanalpha^t \triangleq \forall p \in \Delta^K : \abs{\nsw(p,\mustar)-\nsw(p,\muhat^t)} \le \alpha^t \cdot \sum_{j \in [K]} p_j \cdot r_j^t + \frac{4}{t}.
	\]
	Conditioned on $\cleanmu^t \land \cleanmcd^t$, note that $\cleanalpha^t$ holds for $\alpha^t = N$ due to \Cref{lem:first-bound}, and for $\alpha^t = \sqrt{12 N K \log{NKt}}$ due to \Cref{lem:second-bound}. 
	
	Let $\phat^t \in \argmax_{p \in \Delta^K} \nsw(p,\muhat^t)$. We wish to bound the regret $\nsw(\pstar,\mustar)-\nsw(\phat^t,\mustar)$ that \epsgreedy incurs when performing exploitation in round $t$ by choosing policy $\phat^t$. Conditioned on $\cleanmu^t \land \cleanmcd^t \land \cleanepsgr^t$, we have 
	\begin{align}
	&\nsw(\pstar,\mustar)-\nsw(\phat^t,\mustar)\nonumber\\
	&= \left(\nsw(\pstar,\mustar)-\nsw(\pstar,\muhat^t)\right) + \left(\nsw(\pstar,\muhat^t)-\nsw(\phat^t,\muhat^t)\right) + \left(\nsw(\phat^t,\muhat^t)-\nsw(\phat^t,\mustar)\right)\nonumber\\
	&\le\sum_{p \in \set{\pstar,\phat^t}} \abs{\nsw(p,\mustar)-\nsw(p,\muhat^t)} \le 2 \alpha^t \sqrt{\frac{2 K \log(N K t)}{\gamma \cdot t \epsilon^t}} + \frac{8}{t},\label{eqn:epsgreedy-alpha-bound}
	\end{align}
	where the penultimate transition holds because $\phat^t$ is the optimal policy under $\muhat^t$, so $\nsw(\pstar,\muhat^t) \le \nsw(\phat^t,\muhat^t)$, and the final transition follows from \Cref{eqn:epsgreedy-bound} and the fact that $\cleanmu^t \land \cleanmcd^t$ imply $\cleanalpha^t$.  
	
	We are now ready to analyze the expected regret of \epsgreedy at round $T$. We have
	\begin{align*}
	\E[R^T] &= \sum_{t=1}^T \E[r^t] \le \sum_{t=1}^T \E\br{\epsilon^t \cdot 1 + (1-\epsilon^t) \cdot \left(\nsw(\pstar,\mustar)-\nsw(\phat^t,\mustar)\right)}\\
	&\le \sum_{t=1}^T \Bigg(\epsilon^t + \Pr\br{\cleanmu^t \land \cleanmcd^t \land \cleanepsgr^t} \cdot \E\br{\nsw(\pstar,\mustar)-\nsw(\phat^t,\mustar)\ \Big|\ \cleanmu^t \land \cleanmcd^t \land \cleanalpha^t} \\
	&\quad\qquad\qquad + \Pr\br{\neg\cleanmu^t \lor \neg\cleanmcd^t \lor \neg\cleanepsgr^t} \cdot 1 \Bigg)\\
	&\le \sum_{t=1}^T \left(\epsilon^t +  2 \alpha^t \sqrt{\frac{2 K \log(N K t)}{\gamma \cdot t \epsilon^t}} + \frac{8}{t} + \frac{4}{t^3} + \frac{1}{t} \right),
	\end{align*}
	where the final transition holds due to \Cref{eqn:epsgreedy-alpha-bound}, \Cref{lem:muhat-close}, \Cref{lem:second-bound}, and \Cref{eqn:cleanepsgr}. Notice that we are using the fact that 
	\[
	\Pr[\cleanmu^t \land \cleanmcd^t] =\Pr[\cleanmu^t]\cdot\Pr[\cleanmcd^t | \cleanmu^t] \ge (1-2/t^3)\cdot(1-2/t^3) \ge 1-4/t^3.
	\]
	
	To obtain the first regret bound, we set $\epsilon^t = \bigTh{N^{\frac{2}{3}} K^{\frac{1}{3}} t^{-\frac{1}{3}} \log^{\frac{1}{3}}(N K t)}$ and $\alpha^t = N$, and obtain
	\[
	\E[R^T] = \bigOh{N^{\frac{2}{3}} K^{\frac{1}{3}} \log^{\frac{1}{3}}(N K T) \sum_{t=1}^T t^{-\frac{1}{3}}} = \bigOh{N^{\frac{2}{3}} K^{\frac{1}{3}} T^{\frac{2}{3}} \log^{\frac{1}{3}}(N K T)}.
	\]
	
	For the second regret bound, we set $\epsilon^t = \bigTh{N^{\frac{1}{3}} K^{\frac{2}{3}} t^{-\frac{1}{3}} \log^{\frac{2}{3}}(N K t)}$ and $\alpha^t = \sqrt{12 N K \log(N K t)}$, and obtain
	\[
	\E[R^T] = \bigOh{N^{\frac{1}{3}} K^{\frac{2}{3}} \log^{\frac{2}{3}}(N K T) \sum_{t=1}^T t^{-1/3}} = \bigOh{N^{\frac{1}{3}} K^{\frac{2}{3}} T^{\frac{2}{3}} \log^{\frac{2}{3}}(N K T)}.
	\]
	
	Note that in both cases, we omit the $\bigOh{1/t}$ and $\bigOh{1/t^3}$ terms because they are dominated by the $\bigOh{1/t^{1/3}}$ term. In both cases, we use \Cref{lem:powers} at the end.
\end{proof}

\section{\ucb}\label{sec:ucb}
\begin{algorithm}[htb]
	\setstretch{1.2}
	\DontPrintSemicolon 
	\SetAlgoLined
	\KwIn{Number of agents $N$, number of arms $K$}
	\KwParam{Confidence parameter $\alpha^t$ for each $t \in \bbN$}
	\tcp{Pull each arm once}
	\For{$t = 1,\ldots,K$}{
		$p^t \gets$ policy that puts probability $1$ on arm $t$ \tcp*{Pull arm $t$ deterministically}
	}
	\For{$t = K+1,\ldots$}{
		Compute the estimated reward matrix $\muhat^t$\;
		$p^t \gets \argmax_{p \in \Delta^K} \nsw(p,\muhat^t) + \alpha^t \sum_{j \in [K]} p_j \cdot r_j^t$, where $r_j^t \triangleq \sqrt{\frac{\log(NKt)}{n_j^t}}$.
	}
	\caption{\ucb}
	\label{algo:ucb}
\end{algorithm}

In the classical multi-armed bandit setting, \ucb first pulls each arm once. Afterwards, it merges exploration and exploitation cleverly by pulling, in each round, an arm maximizing the sum of its estimated reward and a confidence interval term similar to $r_j^t$ in \Cref{algo:ucb}. Our multi-agent variant similarly selects a policy that maximizes the estimated Nash social welfare plus a confidence term for a policy, which simply takes a linear combination of the confidence intervals of the arms. 

Unlike \explorefirst and \epsgreedy, which can be implemented efficiently, it is not clear if our \ucb variant admits an efficient implementation due to this step of computing $\argmax_{p \in \Delta^K} \nsw(p,\muhat) + \alpha^t \sum_{j \in [K]} p_j r_j^t$. Due to the added linear term, the objective is no longer log-concave. This remains a challenging open problem. However, we notice that this can also be viewed as the problem of optimizing a polynomial over a simplex, which, while NP-hard in general, is known to admit a PTAS when the degree is a constant~\cite{DLP06,DLS15}. Hence, in our case, when the number of agents $N$ is a constant, this step can be computed approximately, but it remains to be seen how this approximation translates to the final regret bounds. 

We show that \ucb achieves the desired $\sqrt{T}$ dependence on the horizon (up to logarithmic factors). In \Cref{sec:lower-bound}, we show that this is optimal. 

\begin{theorem}\label{thm:ucb}
	\ucb is horizon-independent, and has the following expected regret at any round $T$.
	\begin{itemize}
		\item  If $\alpha^t = N$ for all $t$, $\E[R^T] = \bigOh{N K T^{\frac{1}{2}} \log(NKT)}$.
		\item  If $\alpha^t = \sqrt{12 N K \log(N K t)}$ for all $t$, $\E[R^T] = \bigOh{N^{\frac{1}{2}} K^{\frac{3}{2}} T^{\frac{1}{2}} \log^{\frac{3}{2}}(N K T)}$.
	\end{itemize}
\end{theorem}
\begin{proof}
	Fix one of two parameter choices:
	\begin{enumerate}
		\item $\alpha^t = N$ for all $t$ and $c = N$. 
		\item $\alpha^t = \sqrt{12NK\log(NKt)}$ for all $t$ and $c = \sqrt{12NK\log(NKT)}$. 
	\end{enumerate}
	Note that in both cases, $\alpha^t \le c$ for all $t$. Hence, $c$ serves as an upper bound on $\alpha^t$ that does not depend on $t$. We show that in both cases, running UCB with the $\alpha^t$ parameter value yields a regret bound of $\E[R^T] = O(c K \sqrt{T} \log(NKT))$. Substituting the two choices of $c$ then yields the two regret bounds.  
	
	Let us again focus on the event 
	\[
	\cleanalpha^t \triangleq \forall p \in \Delta^K : \abs{\nsw(p,\mustar)-\nsw(p,\muhat^t)} \le \alpha^t \cdot \sum_{j \in [K]} p_j \cdot r_j^t + \frac{4}{t}.
	\]
	Recall the clean events $\cleanmu^t$ and $\cleanmcd^t$ defined in \Cref{lem:muhat-close,lem:second-bound}. As argued in the proof of \Cref{thm:epsgreedy}, we have that $\cleanmu^t \land \cleanmcd^t$ implies $\cleanalpha^t$ for both choices of $\alpha^t$; for $\alpha^t = N$, it follows from \Cref{lem:first-bound}, and for $\alpha^t = \sqrt{12 N K \log(N K t)}$, it follows from \Cref{lem:second-bound}. Using \Cref{lem:muhat-close,lem:second-bound} as well as the union bound, we have that $\Pr[\neg\cleanalpha^t] \le 1-\Pr[\cleanmu^t \land \cleanmcd^t] = 1-\Pr[\cleanmu^t]\cdot\Pr[\cleanmcd^t | \cleanmu^t] \le 1-(1-2/t^3)\cdot(1-2/t^3) \le 4/t^3$. 
	
	Define a clean event $\cleanalpha^* \triangleq \bigwedge_{t \ge \sqrt{T}} \cleanalpha^t$. Here, we do not care about the first $\sqrt{T}$ rounds because the maximum regret from these rounds is $\bigOh{\sqrt{T}}$, which is permissible given our desired regret bounds. By the union bound, we have $\Pr[\neg\cleanalpha^*] \le T \cdot 4/(\sqrt{T})^3 = 4/\sqrt{T}$. Thus, $\cleanalpha^*$ is a high-probability event. In what follows, we derive an upper bound on the expected regret conditioned on $\cleanalpha^*$, i.e., $\E[R^T | \cleanalpha^*]$. Since conditioning on a high-probability event does not affect the expected value significantly, the desired regret bound will then follow. 
	
	For any $t \in [T]$, conditioned on $\cleanalpha^t$ we have that
	\begin{align*}
	\nsw(\pstar,\mustar) &\le \nsw(\pstar,\muhat^t) + \alpha^t \sum_{j \in [K]} \pstar_j \cdot r_j^t + \frac{4}{t} \\
	&\le \nsw(p^t,\muhat^t) + \alpha^t \sum_{j \in [K]} p^t_j \cdot r_j^t + \frac{4}{t} \\
	&\le \nsw(p^t,\mustar) + 2\alpha^t \sum_{j \in [K]} p^t_j \cdot r_j^t + \frac{8}{t},
	\end{align*}
	where the first and the last transition are from conditioning on $\cleanalpha^t$, and the second transition is because $p = p^t$ maximizes the quantity $\nsw(p,\muhat^t)+\alpha^t \sum_{j \in [K]} p_j \cdot r_j^t$ in the \ucb algorithm. 
	
	Let us write $p^{[T]} = (p^1,\ldots,p^T)$ for the random variable denoting the policies used by the algorithm, and $\bar{p}^{[T]} = (\bar{p}^1,\ldots,\bar{p}^T)$ to denote a specific value in $(\Delta^K)^T$ taken by the random variable. 
	
	Instead of analyzing $\E[R^T | \cleanalpha^*]$ directly, we further condition on UCB choosing a specific sequence of policies $\bar{p}^{[T]}$. That is, we are interested in deriving an upper bound on $\E[R^T | \cleanalpha^* \land p^{[T]} = \bar{p}^{[T]}]$.\footnote{Note that even after conditioning on $p^{[T]} = \bar{p}^{[T]}$, there is still randomness left in sampling actions from the policies and sampling the rewards of those actions.} Interestingly, we show that this quantity is $\bigOh{c K \sqrt{T} \log (N K T)}$ for \emph{every} possible $\bar{p}^{[T]}$. 
	
	Fix an arbitrary $\bar{p}^{[T]}$. For $t \in [T]$ and $j \in [K]$, define $q^t_j = \sum_{s=1}^t \bar{p}^s_j$. Then, $\E[n^t_j | p^{[T]} = \bar{p}^{[T]}] = q^t_j$. For each $j \in [K]$, let $T_j$ be the smallest $t$ for which $q^t_j \ge 2\sqrt{T \log (NKT)}$ (if no such $t$ exists, let $T_j = T$); note that given $\bar{p}^{[T]}$, $T_j$ is fixed and not a random variable. Also, we have that $q^{T_j}_j = \bigTh{\sqrt{T \log (NKT)}}$ for each $j \in [K]$. 
	
	Let us define a clean event $\cleanucb \triangleq \forall j \in [K], n^{T_j}_j \ge \sqrt{T \log (NKT)}$. We first show that this is a high probability event. Indeed, using Hoeffding's inequality, we have that for each $j \in [K]$, 
	\begin{align*}
	\Pr\left[n^{T_j}_j < \sqrt{T \log (NKT)}\ \big|\ \cleanalpha^* \land p^{[T]} = \bar{p}^{[T]}\right] &\le \Pr\left[n^{T_j}_j < s^{T_j}_j - \sqrt{T \log(NKT)}\ \big|\ \cleanalpha^* \land p^{[T]} = \bar{p}^{[T]}\right] \\
	&\le \frac{1}{N^2 K^2 T^2}.
	\end{align*}
	Taking union bound over $j \in [K]$, we have that $\Pr\br{\neg\cleanucb\ \big|\ \cleanalpha^* \land p^{[T]} = \bar{p}^{[T]}} \le \frac{1}{N^2 K T^2}$. 
	
	Next, we bound $\E[R^T | \cleanalpha^* \land p^{[T]} = \bar{p}^{[T]}]$ by using event $\cleanucb$. 
	\begin{align}
	&\E\br{R^T\ \big|\ \cleanalpha^* \land p^{[T]} = \bar{p}^{[T]} } \nonumber\\
	&= \sum_{t=1}^T  \E\br{\nsw(\pstar,\mustar)-\nsw(p^t,\mustar)\ \Big|\ \cleanalpha^* \land p^{[T]} = \bar{p}^{[T]} } \nonumber\\
	&\le \max(K,\sqrt{T}) + \sum_{t=\max(K,\sqrt{T})+1}^T \Bigg(1 \cdot \E\br{\nsw(\pstar,\mustar)-\nsw(p^t,\mustar)\ \big|\ \cleanalpha^* \land p^{[T]} = \bar{p}^{[T]} \land \cleanucb}  \nonumber\\
	&\qquad\qquad\qquad\qquad\qquad\qquad\qquad + \Pr\br{\neg\cleanucb\ \big|\ \cleanalpha^* \land p^{[T]} = \bar{p}^{[T]}} \cdot 1\Bigg)\nonumber\\
	&\le \max(K,\sqrt{T}) + \sum_{t=\max(K,\sqrt{T})+1}^T \E\br{2\alpha^t \sum_{j \in [K]} \bar{p}^t_j \cdot r_j^t + \frac{8}{t}\ \Big|\ \cleanalpha^* \land p^{[T]} = \bar{p}^{[T]} \land \cleanucb}\nonumber\\
	&\qquad\qquad\qquad+ T \cdot \Pr\br{\neg\cleanucb\ \big|\ \cleanalpha^* \land p^{[T]} = \bar{p}^{[T]}}\\
	&\le \max(K,\sqrt{T}) + 1 + 2 c \sqrt{2 \log(N K T)} \sum_{t=1}^T  \sum_{j \in [K]}  \frac{\bar{p}^t_j}{\sqrt{c_j}}.\label{eqn:ucb-main}
	\end{align}
	The final transition holds because $\alpha^t \le c$ for all $t$, $r_j^t = \sqrt{\frac{2\log(NKT)}{n_j^t}}$, and conditioned on $\cleanucb$, $n_j^t \ge c_j$ for each $j \in [K]$ and $t \in [T]$, where $c_j = 1$ if $t < T_j$, and $c_j = \sqrt{T \log(N K T)}$ if $t \ge T_j$. Hence,
	\begin{align*}
	\E\br{R^T\ \Bigg|\ \cleanalpha^* \land p^{[T]} = \bar{p}^{[T]} } 
	&\le \max(K,\sqrt{T}) + 1 + 2 c \sqrt{2\log(N K T)} \sum_{j \in [K]} \sum_{t=1}^T    \frac{\bar{p}^t_j}{\sqrt{c_j}}\\
	&= \max(K,\sqrt{T}) + 1 + 2 c \sqrt{2\log(N K T)} \sum_{j \in [K]} \left( \sum_{t=1}^{T_j-1}    \frac{\bar{p}^t_j}{1} +\sum_{t=T_j}^T    \frac{\bar{p}^t_j}{\sqrt{T \log(NKT)}} \right)\\
	&\le \max(K,\sqrt{T}) + 1 + 2 c\sqrt{2\log(N K T)} \sum_{j \in [K]} \left( q^{T_j}_j +  \frac{T}{\sqrt{T \log(NKT)}} \right)\\
	&= \bigOh{c K \sqrt{T} \log(NKT)}.
	\end{align*}
	
	Because this bound holds for every possible $\bar{p}^{[T]}$, we also have that $\E[R^T | \cleanalpha^*] = \bigOh{c K \sqrt{T} \log(NKT)}$. Finally, we can see that
	\begin{align*}
	\E[R^T] &= \Pr[\cleanalpha^*] \cdot \E[R^T | \cleanalpha^*] + \Pr[\neg\cleanalpha^*] \cdot \E[R^T | \neg\cleanalpha^*]\\
	&\le 1 \cdot \bigOh{c K \sqrt{T} \log(NKT)} + \frac{4}{\sqrt{T}} \cdot 1 = \bigOh{c K \sqrt{T} \log(NKT)}.
	\end{align*}
	Recall that substituting $c = N$ and $c = \sqrt{12 N K \log(N K T)}$ yields the two regret bounds.
\end{proof}

We emphasize that our analysis of the multi-agent UCB differs significantly from the analysis of the classical (single-agent) UCB. For example, the use of clean event $\cleanalpha^*$ is unique to our analysis. More importantly, the expression in \Cref{eqn:ucb-main} is also unique to our setting in which the algorithm can ``pull'' a probability distribution over the arms. The corresponding expression in case of the classical UCB turns out to be much simpler and straightforward to bound. In contrast, we need to use additional tricks to derive the bound of $\bigOh{cK\sqrt{T}\log(NKT)}$.  

Finally, in the proof presented above, we showed that, assuming the clean event $\cleanalpha^*$, the expected regret is small conditioned on \emph{any} sequence of policies that the UCB algorithm might use. At the first glance, this may seem surprising. However, a keen reader can observe that the clean event $\cleanalpha^*$ can only occur when the UCB algorithm uses a ``good'' sequence of policies that leads to low expected regret. A similar phenomenon is observed in the analysis of the classical (single-agent) UCB algorithm as well (see, e.g.,~\cite{Sliv19}): assuming a different clean event, the classical UCB algorithm is guaranteed to not pull suboptimal arms too many times. 

\section{Lower Bound}\label{sec:lower-bound}
We lastly turn our focus to proving lower bounds on the expected regret of any algorithm for our multi-agent multi-armed bandit (MA-MAB) problem. In the classical multi-armed bandit problem, it is known that no algorithm can achieve a regret bound of $E[R^T] = o(\sqrt{KT})$, when the constant inside the little-Oh notation is required to be independent  of the distributions in the given instance~\cite{ACFS02}. For further discussion on bounds where the constant is allowed to depend on the distributions in the given instance, we refer the reader to \Cref{sec:disc}. Our goal in this section is to reproduce this lower bound for our multi-agent multi-armed bandit problem. This would establish that the $\sqrt{T}$-dependence of the expected regret of our UCB variant on the horizon $T$ from \Cref{thm:ucb} is optimal. Note that our focus is solely on the dependence of the expected regret on $T$ as $T$ is typically much larger than both the number of agents $N$ and the number of arms $K$. We leave it to future work to optimize the dependence on $N$ and $K$. 

First, we notice that any lower bound derived for the case of a single agent also holds when there are $N > 1$ agents. This is because one can consider instances in which all but one of the agents derive a fixed reward of $1$ from every arm. Note that the contribution of such agents to the product in the Nash social welfare expression is always $1$ regardless of the policy chosen. Hence, the Nash social welfare reduces to simply the expected utility of the remaining agent, i.e., the Nash social welfare in an instance with only this one agent. Therefore, any lower bound on the expected regret that holds for MA-MAB with a single agent also holds for MA-MAB with $N > 1$ agents. 

Next, let us focus on the MA-MAB problem with $N=1$ agent. At the first glance, this may look almost identical to the classical multi-armed bandit problem. After all, if there is but one agent, the policy maximizing the Nash social welfare places probability $1$ on the arm $j^*$ that gives the highest mean reward to the agent. Thus, like in the classical problem, our goal would be to converge to pulling arm $j^*$ repeatedly and our regret would also be measured with respect to the best policy which deterministically pulls arm $j^*$ in every round. However, there are two subtle differences which prevent us from directly borrowing the classical lower bound.
\begin{enumerate}
	\item In our MA-MAB problem, an algorithm is allowed to ``pull'' a \emph{distribution over the arms} $p^t$ in round $t$ and learn the stochastically generated rewards for a \emph{random} arm $j^t$ sampled from this distribution. This makes the algorithm slightly more powerful than an algorithm in the classical MAB problem which must deterministically choose an arm to pull. 
	\item In our MA-MAB problem, the regret in round $t$ is computed as the difference between the mean reward of the best arm and the \emph{expected} mean reward of an arm $j^t$ sampled according to the distribution $p^t$ used by the algorithm. In the classical problem, one would replace the latter term with the mean reward of the arm actually pulled in round $t$. 
\end{enumerate}

The latter distinction is not particularly troublesome because our focus is on the \emph{expected} regret of an algorithm anyway. However, the first distinction makes it impossible to directly borrow lower bounds from the classical MAB problem. 

One might wonder if there is still a way to reduce the MA-MAB problem with $N=1$ agent to the classical MAB problem. For example, given an algorithm $A$ for MA-MAB with $N=1$, what if we construct an algorithm $\hat{A}$ for the classical MAB and use the lower bound on the expected regret of $\hat{A}$ to derive a lower bound on the expected regret of $A$? The problem with such reduction is that once $A$ chooses a distribution $p^t$, we have no control over which arm will be sampled. This choice is crucial as it will determine what information the algorithm gets to learn. We cannot mimic this learning process in our deterministic algorithm $\hat{A}$. Upon careful consideration, it also seems difficult to express the expected regret of $A$ as the convex combination of the expected regret of several deterministic algorithms for the classical MAB. 

Instead of aiming to find a black-box reduction to the classical problem, we therefore investigate in detail the proof of the $\bigOm{\sqrt{KT}}$ lower bound for the classical MAB due to \citet[Theorem~5.1]{ACFS02} and observe that their argument goes through for our MA-MAB problem as well. Instead of repeating their proof, we survey the key steps of their proof in which they assume the algorithm to be deterministically pulling an arm and highlight why the argument holds even when this is not the case. 
\begin{itemize}
	\item In the proof of their Lemma~A.1, in the explanation of their Equation~(30), they cite the assumption that given the rewards observed in the first $t-1$ rounds (they denote this by the vector $\mathbf{r}^{t-1}$), the algorithm pulls a fixed arm $i_t$ in round $t$. They refer to the distribution $\mathbf{P}_i\{r^t | \mathbf{r}^{t-1}\}$ of the reward in round $t$ given $\mathbf{r}^{t-1}$. In their case, the randomness in $r^t$ is solely due to stochasticity of the rewards since the arm pulled ($i_t$) is fixed. However, in our case, one can think of $\mathbf{P}_i\{r^t | \mathbf{r}^{t-1}\}$ as containing randomness both due to the random choice of $i_t$ and due to the stochasticity of the rewards, and their equations still go through. 
	\item In the same equation, they consider $\mathbf{P}_{unif}\{i_t = i\}$, the probability that arm $i$ is pulled in round $t$. In their case, the only randomness is due to $\mathbf{r}^{t-1}$. In our case, there is additional randomness due to the sampling of an arm in round $t$ from a distribution $p^t$. However, this does not affect their calculations.
	\item Finally, in the proof of their Theorem A.2, they again consider the probability $\mathbf{P}_i\{i_t=i\}$ and the same argument as above ensures that their proof continues to hold in our setting. 
\end{itemize}

Thus, we have the following lower bound. 
\begin{proposition}\label{thm:lower-bound}
	For any algorithm for the MA-MAB problem, there exists a problem instance such that $\E[R^T] = \bigOm{\sqrt{KT}}$.
\end{proposition}

\section{Discussion}\label{sec:disc}
We introduce a multi-agent variant of the multi-armed bandit problem in which different agents have different preferences over the arms and we seek to learn a tradeoff between the arms that is fair with respect to the agents, where the Nash social welfare is used as the fairness notion. Our work leaves several open questions and directions for future work. 

\emph{Computation.} As we observed in the paper, our \explorefirst and \epsgreedy variants can be implemented in polynomial time. However, for our UCB variant, it is not clear if the step of computing $\argmax_{p \in \Delta^K} \nsw(p,\muhat) + \alpha^t \sum_{j \in [K]} p_j r_j^t$ can be computed in polynomial time due to the added linear term at the end. As we mentioned in \Cref{sec:ucb}, there exists a PTAS for this step when the number of agents $N$ is constant, but the complexity in the general case remains open. One might wonder if it helps to take the logarithm of the Nash social welfare, i.e., solve $\argmax_{p \in \Delta^K} \log\nsw(p,\muhat) + \alpha^t \sum_{j \in [K]} p_j r_j^t$. Indeed, since $\log\nsw$ is a concave function, this can be solved efficiently. However, our key lemmas use bounds on the $\nsw$ function that do not hold for $\log\nsw$ function. Further, such an approach would yield a regret bound where the regret is in terms of $\log\nsw$, which cannot be easily converted into regret in terms of $\nsw$. 

\emph{Logarithmic regret bound for \ucb.} In the classical stochastic multi-armed bandit setting, \ucb has two known regret bounds with optimal dependence on $T$. There is an \emph{instance-independent} bound that grows roughly as $\sqrt{T}$ (where the constants depend only on $K$ and not on the unknown mean rewards in the given instance) and an \emph{instance-dependent} bound that grows roughly as $\log T$ (where the constants may depend on the unknown mean rewards in the given instance in addition to $K$). While we recover the former bound in our multi-agent setting, we were not able to derive an instance-dependent logarithmic regret bound. This remains a major challenging open problem. 

\emph{Improved lower bounds.} In \Cref{sec:lower-bound}, we observe that the instance-independent $\Omega(\sqrt{KT})$ lower bound from the classical setting also holds in our multi-agent setting. Given that our upper bounds increase with $N$, it would be interesting to see if we can derive lower bounds that also increase with $N$. Deriving instance-dependent lower bounds in our setting would also be interesting. 

\emph{Fairness.} While maximizing the Nash social welfare is often seen as a fairness guarantee of its own, as discussed in the introduction, the policy with the highest Nash social welfare is also known to satisfy other fairness guarantees. However, it is not clear if the additive regret bounds we derive in terms of the Nash social welfare also translate to bounds on the amount by which these other fairness guarantees are violated. Considering other fairness guarantees and bounding their total violation is also an interesting direction for the future. 

\emph{Multi-agent extensions.} More broadly, our work opens up the possibility of designing multi-agent extensions of other multi-armed bandit problems. For example, one can consider a multi-agent dueling bandit problem, in which an algorithm asks an agent (or all agents) to compare two arms rather than report their reward for a single arm. Meaningfully defining the regret for such frameworks and designing algorithms that bound it is an exciting future direction. 

\bibliographystyle{plainnat}
\bibliography{abb,bibliography}
\end{document}